\documentclass[10pt,letterpaper,conference]{IEEEtran}
\usepackage[dvips]{graphicx}
\usepackage{amssymb,amsmath}
\usepackage{textcomp}
\usepackage{url}

\usepackage[absolute]{textpos}

\TPMargin{5pt}

\newcommand{\copyrightstatement}{
   \begin{textblock}{0.8}(0.10,0.95)    
        \noindent
        \footnotesize
        G. B. T. Kalejaiye, et al. ``Mobile Offloading in Wireless Ad Hoc
        Networks'', in E. Baccelli, F. Juraschek, O. Hahm, T. C. Schmidt,
        H. Will, M.~W\"ahlisch (Eds.), Proc. of 3rd MANIAC Challenge, Berlin, Germany, July 27 - 28, 2013,  arXiv, Jan. 2014.
        
   \end{textblock}
}

\begin{document}
\title{Mobile Offloading in Wireless Ad Hoc Networks}
\author{
    \authorblockN{Gabriel B. T. Kalejaiye \ \ \ \ Jo\~ao A. S. R. Rondina \ \ \ \ Leonardo V. V. L. Albuquerque \ \ \ \ Ta\'is L. Pereira \\ Luiz F. O. Campos \ \ \ \  Raphael A. S. Melo \ \ \ \ Daniel S. Mascarenhas \ \ \ \ Marcelo M. Carvalho\authorrefmark{1}}
    \authorblockA{
        \\ Department of Electrical Engineering, University of Bras\'ilia, Brazil}
    \authorblockA{mmcarvalho@ene.unb.br \authorrefmark{1}}
}


\maketitle

\begin{abstract}

This paper describes a strategy that was designed, implemented, and presented at the {\it Mobile Ad Hoc Networking Interoperability and Cooperation (MANIAC) Challenge 2013}. The theme of the MANIAC Challenge 2013 was ``Mobile Data Offloading,'' and consisted on developing and comparatively evaluating strategies to offload infrastructure access points via customer ad hoc forwarding using handheld devices (e.g., tablets and smartphones). According to the challenge rules, a hop-by-hop bidding contest (or ``auction'') should decide the path of each data packet towards its destination. Consequently, each team should rely on other teams' willingness to forward packets for them in order to get their traffic across the network. In this application scenario, the incentive for customers to join the ad hoc network is discounted monthly fees, while operators should benefit from decreased infrastructure costs. Following the rules of MANIAC Challenge 2013, this paper proposes a strategy that is based on the concept of how ``tight'' a node is to successfully deliver a packet to its destination within a given deadline. This ``tightness'' idea relies on a shortest-path analysis of the underlying network graph, and it is used to define three sub-strategies that specify a) how to participate in an auction; b) how to announce an auction; and c) how to decide who wins the announced auction. The proposed strategy seeks to minimize network resource utilization and to promote cooperative behavior among participant nodes. 

\end{abstract}

\section{Introduction}
\label{sec:introduction}

\copyrightstatement

The advent of mobile devices such as smartphones and tablets have posed an unprecedent traffic demand on current mobile communications infrastructure. For instance, AT\&T has reported that wireless data traffic has increased by 20,000\% in just five years (2007-2012), and that an exponential traffic increase should be expected in the years to come~\cite{att}. Following similar reports, a number of operators have started to look for alternative solutions to alleviate the impact of the so-called ``mobile data crunch'' problem. As an immediate (and partial) solution to this problem, many mobile operators are currently deploying their own WiFi networks for offloading their core network infrastructure by encouraging their customers to switch to WiFi as much as possible. 

Motivated by this timely and important issue, the organizers of the {\it Mobile Ad Hoc Networking Interoperability and Cooperation (MANIAC) Challenge 2013}~\cite{maniac.challenge.2013} have called for the development and comparative evaluation of strategies to offload infrastructure access points via customer ad hoc forwarding using handheld devices (e.g., smartphones and tablets). In this year's challenge, competing teams should come together to form a wireless ad hoc network while simultaneously connected to a backbone of access points. Then, traffic from the backbone would be generated and destined to some other backbone node. A hop-by-hop {\it bidding contest} should decide the path of each data packet towards its destination, and each team should rely on other teams' willingness to forward traffic for them in order to get their traffic across the network. In this application scenario, the incentive for customers to participate in the ad hoc network is discounted monthly fees, while operators would benefit from decreased infrastructure costs. Ultimately, the main goal of the MANIAC Challenge 2013 was to demonstrate scenarios/strategies that do not degrade user experience while offering significant mobile data offloading on the infrastructure.


This paper presents a strategy that was designed, implemented, and presented at the MANIAC Challenge 2013. The main idea behind the proposed strategy is related to the concept of how ``tight'' a node is to successfully deliver a packet to its target destination within a given deadline. This ``tightness'' concept is built upon a shortest-path analysis of the underlying network graph, and it is used to define three sub-strategies for participation in the ad hoc network: the {\it bidding} strategy, which sets the bid for an auction based on the budget, fine, timeout, and target destination announced by the auctioneer; the {\it budget-and-fine} setup strategy, which sets the budget and fine for every auction that is announced (in case we need to forward a packet to someone else); and the {\it request-for-bids} strategy, which decides who wins our auctions based on received bids. 

In addition, we have included another feature to the strategy based on observed competitors' behavior during the MANIAC Challenge 2013. This additional feature seeks to promote {\it cooperation} within the network by penalizing selfish or greedy behavior. In fact, in order to convince other clients to join the ad hoc network and share their handheld resources (bandwidth, battery life, etc.), it is fundamental that every participant node perceives some level of fairness and profit sharing within the network in the long term. Otherwise, clients may not feel encouraged to participate in the ad hoc network and, as a result, they may decide to leave it or never join it. Such an effect may compromise the feasibility of the proposed offloading architecture due to severe network partitions and/or poor geographical coverage. 

The paper is divided as follows. Section~\ref{sec:rules} describes the rules and setup of the MANIAC Challenge 2013. The proposed strategy is described in Section~\ref{sec:strategy}, and Section~\ref{sec:lessons_learned} contais a discussion about lessons we have learned from participation in the challenge. Finally, Section~\ref{sec:conclusions} contains the conclusions.  

\section{MANIAC Challenge 2013: Rules \& Setup}
\label{sec:rules}

\begin{figure}[htb]
\centering
\includegraphics[scale=0.6]{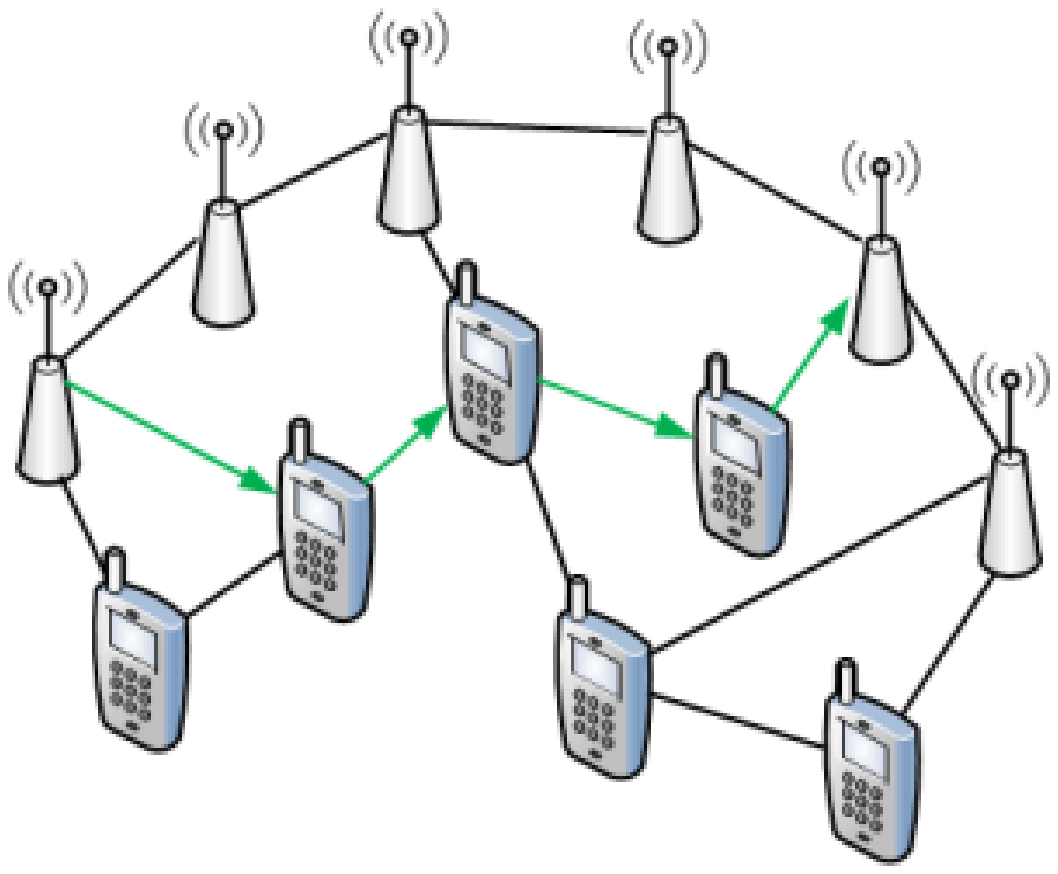}
\caption{MANIAC Challenge 2013 scenario: a wired backbone of WiFi access points  and a collection of mobile handheld devices with backbone coverage.}
\label{fig:scenario}
\end{figure}

The general goal of the MANIAC Challenge is to better understand cooperation and interoperability in ad hoc networks. Therefore, the main idea of the challenge is to put together competing teams of students/researchers to form a wireless ad hoc network, while simultaneously connected to a backbone of access points. Figure~\ref{fig:scenario} depicts a typical scenario, where the lines connecting the devices indicate wireless connectivity, and the green arrows indicate a path that would be taken by a packet to traverse the ad hoc network from the AP source to a target AP destination.  

In the specific case of the MANIAC Challenge 2013, the network setup consisted in the interconnection of the following elements:
\begin{enumerate}
\item A wired backbone of WiFi access points (APs);
\item A collection of Google Nexus 7 tablets for the ad hoc network. 
\end{enumerate}

Each team could compete with two tablets and, because the tablets were Google Nexus~7, all strategies had to be written for the Android Mobile OS. To help with strategy implementation, the organizers have also made available the ``MANIAC framework,'' which included an API that provided function calls for ``bidding''  and ``auction,'' for sending and receiving data, and for accessing topology and routing information through the Optimized Link-State Routing (OLSR) protocol~\cite{clausen2003olsr, olsr}. In particular, the teams were allowed to override the forwarding decisions made by the OLSR routing protocol in order to implement their forwarding strategies in the hop-by-hop bidding contests. 

The competition was based on multiple rounds, where each round consisted of multiple games. After each round, all teams were allowed to refine their strategy. The organizers monitored every packet launched to the network, and collected statistical data to analyze each game and to identify nodes that did not comply with the rules. Two winning categories were defined for the challenge: {\it performance}, based on both maximum balance above zero and highest packet delivery ratio, and {\it design}, where a subjective evaluation would select the most inspiring idea. In the following, we detail the rules that were laid out for this year's challenge.


\subsection{Bidding}

Each game starts with a randomly selected backbone access point (AP) initiating a forwarding request to deliver one data packet to another randomly selected AP. In each bidding request, the source AP indicates:
\begin{itemize}
\item Maximum {\it budget} available for the successful delivery of the data packet;
\item Maximum packet {\it delivery time} (i.e., ``timeout'') translated to a {\it maximum number of hops} that the particular packet could traverse; 
\item A {\it fine} to be paid in case of unsuccessful packet delivery to its destination.  
\end{itemize}

All neighbors that are able to overhear a bidding request are obligated to bid their price lower or equal than the announced budget to forward the data packet. All bids are open, in the sense that they can be detected and understood by anyone within range of the node that sent the bid. After sending a bidding request, every auctioneer must wait for 3 s before broadcasting the auction winner. The initial upstream node (i.e., the AP source) always selects the bid with the lowest cost. From this point on, per packet hop-by-hop recursive auctions should be performed in order to decide each data packet's next hop towards destination. Each handheld device must deliver the packet to the destination either via the ad hoc network or the provider's infrastructure backbone based on independent bidding. Using the backbone for delivery guarantees a 100\% packet delivery ratio. Each backbone node is reachable within radio range of at least one handheld, and the handheld topology is not partitioned. Handheld devices are free to advertise their own maximum budget and fine in their bidding requests, except for the timeout, which has to be the same as the one set by the source AP. However, in all auctions, the advertised fine should be smaller or equal to the fine agreed upon for the previous hop. After receiving the bids, the handheld device can choose a winner downstream node based on its own strategy. A node that wins an auction is allowed to drop the packet it has just received based on its own strategy. In order to avoid loops, a handheld device is not allowed to bid for a data packet it has already forwarded once.

\subsection{Payment}

An upstream node pays the accepted price to the chosen downstream node if the packet has been delivered successfully to the final destination.
If the data packet does not make it to the backbone destination before the timeout, the chosen downstream node must pay the agreed fine to the upstream node.
As mentioned before, the initial data packet budget, the associated fine, and timeout are predefined, per packet, by the organizers. In all auctions, the fine must be smaller or equal to the budget defined in the bidding request. A handheld's balance may be temporarily negative. A handheld that bypass the ad hoc network by using the backbone must pay a price equal to the initial maximum data packet budget (when the data packet was first introduced in the network). The fine has to be paid in case of unsuccessful packet delivery (i.e., packet loss or exceeding the packet delivery time). 

\section{The Proposed Strategy}
\label{sec:strategy}

In this section, we define the proposed strategy, which comprises three sub-strategies: the {\it bidding} strategy, the {\it budget-and-fine} setup strategy, and the {\it request-for-bids} setup strategy. Before going into details about each sub-strategy, we explain the nomenclature and definitions used in this work. 

\subsection{Nomenclature and Definitions}

When the {\it source} access point (AP) announces its {\it request for bids} (RFB), it announces its budget $B_0$, along with a fine $F_0$ to be paid in case the data packet is not delivered to the destination AP after traversing a {\it maximum number} $H_0$ of  hops (i.e., this is the ``timeout'' announced by the source AP, as mentioned before). Let $hc_i$ denote the number of hops (or ``hop count'') of the {\it shortest path} (in terms of number of hops) computed from node $i$ to the {\it destination} AP. Also, let $p_k$ denote the number of hops traversed by a packet from the source AP to a given node $k$ in the network. 


A key metric in our proposal is the definition of a ``tightness function'' $\Delta_i$ for a node $i$ in the network, i.e., $\Delta_i$ measures how ``tight'' a node $i$ is with respect to making the deadline $H_0$ imposed by the source AP.  More specifically, given the timeout $H_0$ announced by the source AP, and the number $p_u$ of hops  already traversed by the data packet all the way to node $i$'s upstream node $u$ (i.e., node $i$'s predecessor, the one who will issue an RFB), $\Delta_i$ measures the ``surplus'' or ``deficit'' (in number of hops) that node $i$ possess with respect to timeout $H_0$ {\it if the data packet were forwarded through its {\it shortest path} to the destination AP}, given by 
\begin{equation}
\Delta_i = (H_0 - p_u - 1) - hc_i, \quad \forall i \in \mathcal{N}(u),
\label{eq:tightness-function}
\end{equation}   
where $\mathcal{N}(u)$ is the set of nodes who are able to overhear the RFB from node $u$, i.e., the neighbors of node $u$. The computation of $\Delta_i$ requires the knowledge of the underlying network topology (the OLSR routing protocol provides that), and the use of any shortest path algorithm. Notice that, if $\Delta_i < 0$, node $i$ cannot deliver the data packet within the deadline (even if the data packet follows node $i$'s shortest path to the destination AP). On the other hand, if $\Delta_i = 0$, node $i$ needs {\it exactly} the number of hops contained in its shortest path to the destinatio AP in order to make the deadline. This is a ``tight'' situation for node $i$, since it relies on the unpredicted outcome of other downstream auctions for the packet to arrive within the deadline. Finally, if $\Delta_i > 0$, node $i$ has a higher chance to deliver the data packet within the deadline because the packet may even deviate from its shortest path to the destination AP, but it has a ``surplus'' of hops before the deadline is up. 

\subsection{The Bidding Strategy}

The bidding strategy defines how the value of the bid is set once an RFB is overheard from an upstream node $u$. For that, we first need to determine the set $\mathcal{N}(u)$ of neighbors of the upstream node $u$. This set contains {\it our competitors} in the upcoming auction, and it can be easily found because all nodes have complete knowledge of the network topology. For each node $i \in \mathcal{N}(u)$, we compute $\Delta_i$ according to Eq.~(\ref{eq:tightness-function}). Based on the values of $\Delta_i$, we create a subset $\mathcal{S}(u) \subseteq \mathcal{N}(u)$ that contains all nodes in $\mathcal{N}(u)$ such that $\Delta_i \ge 0$, i.e., the set $\mathcal{S}(u)$ contains all nodes that are actually able to deliver the packet within the deadline and, therefore, they are the ones most likely to win the auction announced by node $u$ (our actual competitors). Observe that, we are assuming that node $u$ will usually prefer not to pay a fine. 

Given $\mathcal{S}(u)$, we want to estimate how competitive we are in terms of packet delivery from the point of view of node $u$. It is reasonable to expect that the likelihood of successfully delivering a packet will play a key role in any decision making by any node. Therefore, we choose to find out how competitive we are by using our ``tightness function.'' Specifically, we compute how ``tight'' we are with respect to the {\it average tightness} $\overline{\Delta}$ of nodes in $\mathcal{S}(u)$, defined as 
\begin{align}
\overline{\Delta} & = \frac{1}{|\mathcal{S}(u)|}\sum_{i \in \mathcal{S}(u)} (H_0 - p_u - 1) - hc_i \nonumber \\
& = (H_0 - p_u - 1) - \overline{hc},
\end{align}
where $|\mathcal{S}(u)|$ is the cardinality of $\mathcal{S}(u)$, and $\overline{hc}$ is the average optimal hop count over all $i \in \mathcal{S}(u)$, i.e., the average {\it shortest path} to the destination AP computed for each node $i \in \mathcal{S}(u)$. Once the average tightness $\overline{\Delta}$ is found, we can compute our {\it relative tightness} $c_n$ with respect to $\overline{\Delta}$ by 
\begin{equation}
c_n = \frac{\Delta_n}{\overline{\Delta}} = \frac{(H_0 - p_u - 1) - hc_n}{(H_0 - p_u - 1) - \overline{hc}},
\end{equation} 
where the subscript $n$ is used to identify ourselves. It is important to mention that the above computation will only happen if our tightness function is such that $\Delta_n > 0$ and $|\mathcal{S}(u)| > 0$. Otherwise, we have specific rules for making our bid (explained later). 

Observe that, if $c_n < 1$ and $\Delta_n > 0$, our competitors are better positioned than us (on average, with respect to a surplus of hop counts). Therefore, there is a high chance that they become more aggressive to win the bidding, since they may feel that they can deliver the packet in time. At the same time, since $c_n < 1$, it means that we are running a higher risk on not having the packet delivered to its final destination, compared to others. Therefore, we may want to set a higher bid (closer to the budget $B_u$) because the risk should not be worth it to take. In case $c_n \approx 1$, we have similar conditions than other competitors and, therefore, we should try to win the auction with a lower bid compared to previous case. However, if $c_n > 1$, it means that we are better positioned than the average of our competitors. Therefore, we should strive to win the bid by offering a very attractive price (closer to the fine $F_u$).

In additon to $c_n$, another important metric to take into account is how $\Delta_n$ (the value of our tightness function) compares to the {\it highest} value of $\Delta_i$ for $i \in \mathcal{S}(u)$. This is because, if $\Delta_n > \max_{i \in \mathcal{S}(u)}{\Delta_i} = \Delta_{\max}$, it means that we are the best choice for the upstream node $u$ in terms of a positive surplus of hop counts towards destination. Therefore, we should strive to win the auction by becoming as aggressive as possible in our bid (i.e., to set lower values for the bid to make sure we win the auction). Otherwise, if $\Delta_n < < \Delta_{\max}$, we should have very low expectations to win the auction and, therefore, we should not make dramatic changes in our bid for different values of $c_n$. Based on that, we define the parameter $a_n$ that compares our tightness value with the best tightness value in $\mathcal{S}(u)$, i.e.,  
\begin{equation}
a_n = \frac{\Delta_n}{\Delta_{\max}}.
\end{equation}   

Given the values of the budget $B_u$ and the fine $F_u$ announced by the upstream node $u$, and since $F_u \le B_u$ (according to auction rules), our offered bid $O(c_n)$, will be given by a {\it logistic function} of the form 
\begin{equation}
O(c_n) = (B_u - F_u) \left[ 1 - \frac{1}{1 + e^{-a_n(c_n - 1)}} \right] + F_u,
\label{eq:offered_bid}
\end{equation}
where $F_u \le O(c_n) \le B_u$, i.e., we opt for never making a bid less than the established fine $F_u$. As it can be seen, the logistic function is centered on $c_n = 1$, and the steepness of the curve is controlled by $a_n$.

\begin{figure}[htb]
\centering
\includegraphics[scale=0.61]{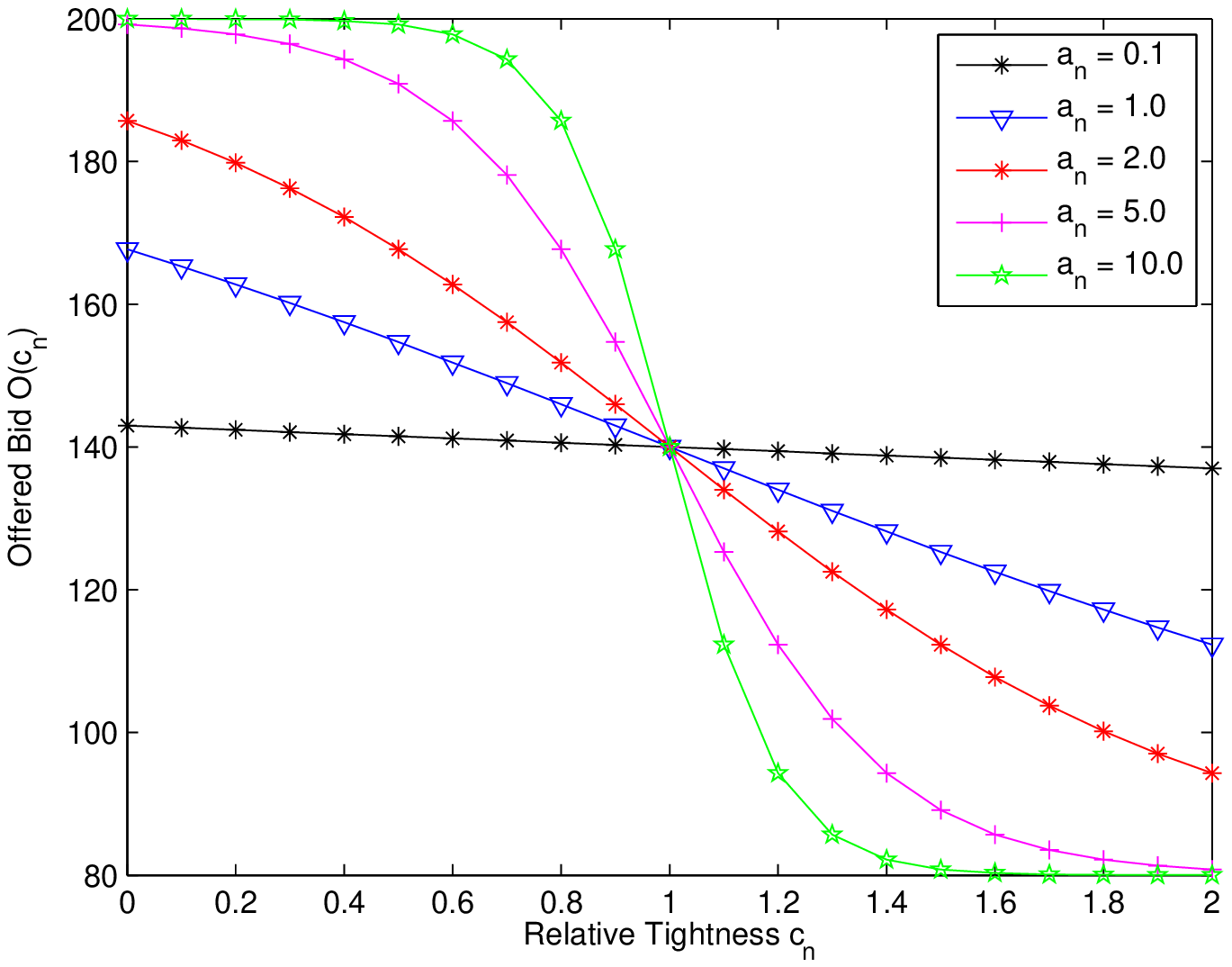}
\caption{Example of offered bid curves $O(c_n)$ for different values of the parameter $a_n$ when $B_u = 200$ and $F_u = 80$.}
\label{fig:offered-bid}
\end{figure}

Finally, if $\Delta_n < 0$, we discourage the upstream node from choosing us by setting our bid equal to the budget $B_u$. Likewise, if there is no competition, i.e., we are the only node reachable by the upstream node, we set our bid to the maximum value $B_u$, and if $\Delta_n = 0$, it means that we are very ``tight'' and, therefore, we should set our bid to $B_u$ (high risk). Figure~\ref{fig:offered-bid} shows examples of offered bid curves for different valus of $a_n$ and $c_n$.  


Lastly, we define a {\it random instant of time} $t_s$ in which we transmit our (mandatory) bid for a given advertised RFB. This is because, according to the auction's rules, there is an {\it auction timeout} ($AT$) before which everyone who gets an RFB must (mandatorily) transmit its bid. Hence, in order to diminish the chances of having competing nodes that always hear our bids before formulating their own ones (e.g., ``sniper-type'' of behavior at the closing of an auction), we select a random instant of time $t_s$ that is uniformly chosen in the interval $[0.5 AT, \ 0.75 AT]$ to transmit our bid.

\subsection{Budget-and-Fine Setup Strategy}

Once an auction is won, the strategy to set the budget $B_n$ and fine $F_n$ to be announced on an RFB is based on a fixed rule. Given that the upstream node paid us an amount equal to our winner offer $O^*$, the budget $B_n$ and fine $F_n$ will be set to 
\begin{equation}
B_n = 0.6 \times O^* \quad \text{ and } \quad F_n = 0.9 \times B_n,
\end{equation} 
where the value of 60\% and 90\% were estimated to be reasonable values that can afford, from time to time, some loss not so high (according to auction rules). 

\subsection{Request-for-Bids Setup Strategy} 

\begin{figure}[htb]
\centering
\includegraphics[scale=0.61]{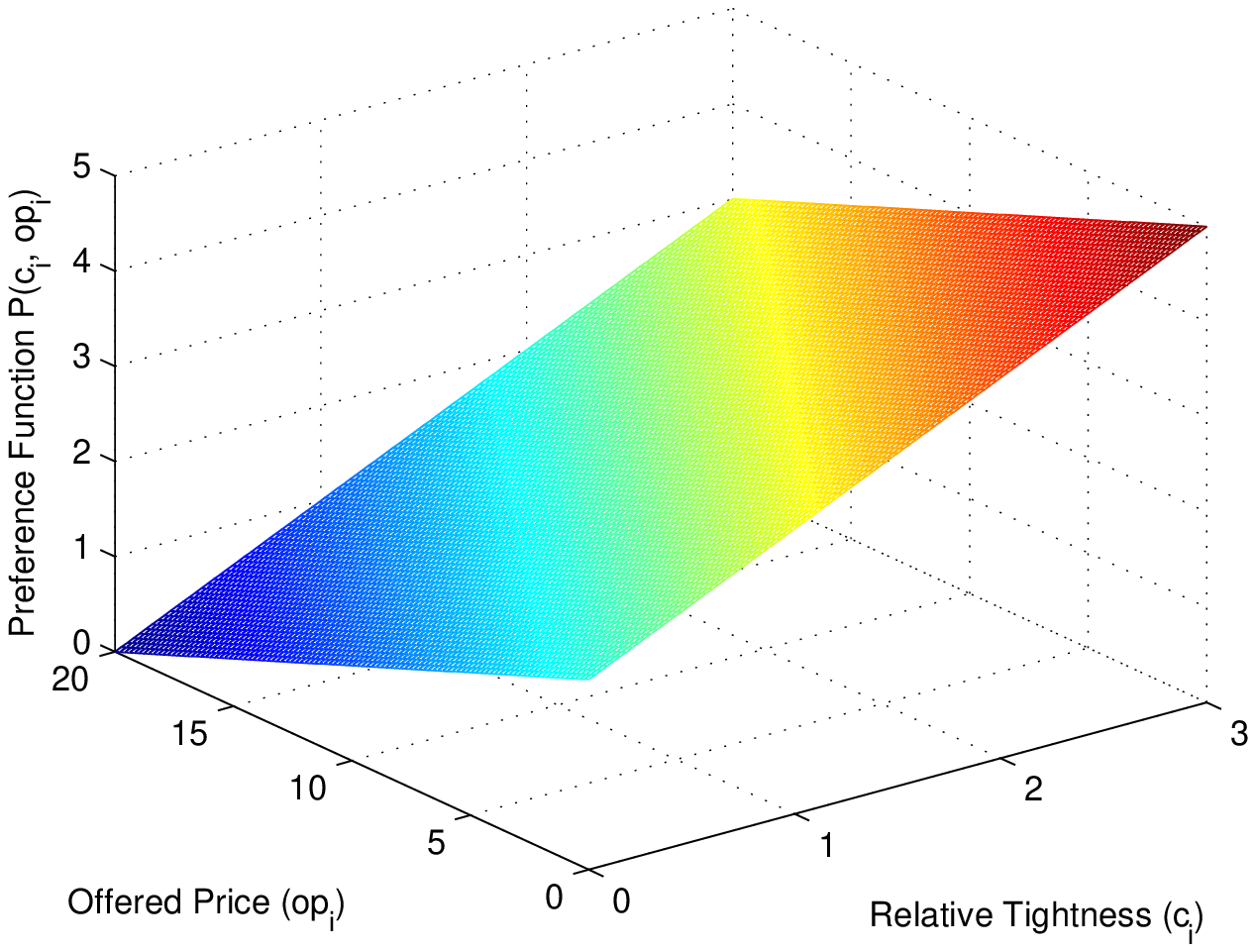}
\caption{Example of preference function for the values $B_n = 20$, $c_{\max} = 3$, $k_1 = 2$, and $k_2 = 3$}
\label{fig:plano}
\end{figure}

In order to determine who wins our RFB, we want to consider both the offered price $op_i$ and the {\it relative tightness} $c_i$ of each neighbor $i$ around us (except for the upstream node from whom we received the packet). Based on those values, we want to determine the best neighbor to foward our packet to. The maximum offered price we can receive is our own budget $B_n$, and the maximum relative tightness $c_{\max}$ is related to the neighbor with the highest tightness value $\Delta_i$. In our strategy, we value a high relative tightness more than a good offered price, since we want to guarantee packet delivery as much as possible. Therefore, we need to define a {\it preference function} $P_n(c_i, op_i)$ that translates our preference towards the received bids. Hence, the lowest preference would be given to the node that has $c_i = 0$ and $op_i = B_n$, i.e., $P_n(0, B_n) = 0$. On the hand, the highest preference would be given to the one where $c_i = c_{\max}$ and $op_i = 0$ (for free). Other interesting cases are $P_n(0, 0)$, where it reflects the case when the node is ``tight,'' but it has offered ``free forward'', and the case $P_n(B_n, c_{\max})$ where the node has offered the maximum bid, but it is the best node. If $P_n(0, 0) = k_1$ and $P_n(B_n, c_{\max}) = k_2$, we may set $k_2 > k_1 > 0$ to reflect our tendency to favor packet delivery as opposed to earn money. The absolute values $k_1$ and $k_2$ are not important, but their relative magnitudes should translate our relative preference between those scenarios (adjustable according to the game). The equation of a {\it plane} that intersects those points can be obtained so that the {\it preference function} $P_n(c_i, op_i)$ is finally defined as    
\begin{equation}
P_n(op_i, c_i) = k_1 - \left(\frac{k_1}{B_n}\right) op_i + \left(\frac{k_2}{c_{\max}}\right) c_i 
\end{equation}
Figure~\ref{fig:plano} depicts the case for the hypotetical values $B_n = 20$, $c_{\max} = 3$, and $k_1 = 2$, and $k_2 = 3$.  

\subsection{Strategy Extension: Promoting Cooperative Behavior}
\label{sec:cooperative_behavior}

The original strategy was supplemented with an additional feature as a result of observed competitors' behavior during the MANIAC Challenge event. In fact, considering that {\it i}) performance criteria was based on packet delivery ratio and maximum balance above zero, and {\it ii}) challenge rules enforced the sending of a bid by every node that could hear an RFB, it was possible to have a situation where the auctioneer could actually announce an RFB with {\it zero} budget and, yet, to have his packet disputed by every node in its neigborhood. Hence, in spite of not providing any monetary incentives for other nodes to forward his packet, every single neighbor was forced to participate in a zero-budget auction because of competition rules. Thus, with a zero-budget auction, the auctioneer could make a significant profit based on {\it free} work provided by downstream nodes. For instance, the auctioneer could retain all budget he might have received from an upstream node while, at the same time, to have his packet delivered by downstream nodes for free.

Considering that one of the main goals of MANIAC Challenge 2013 was to find ways of promoting mobile data offloading via customer ad hoc forwarding using handheld devices, and given that customers may not be willing to share their resources if no incentives are given, greedy strategies such as this may actually work {\it against} the feasibility of ad hoc networks as a solution to the offloading problem. In fact, if a significant portion of participant nodes adopt greedy strategies, other nodes may realize that it is not worth it to forward someone's else traffic. Consequently, many nodes may leave the ad hoc network, which may cause severe network partitions and compromise the own existence of the network. Therefore, it is important to {\it discourage} anyone from performing such practices by signaling, somehow, that such strategies are not acceptable and they are hurtful to the well-being of the network as a whole. Instead, a {\it cooperative behavior} should be promoted among nodes, by allowing everyone to obtain some benefit out of their participation in the ad hoc network. 

With that goal in mind, we have added another rule as a follow-up to our bidding strategy: it states that if a request for bids with zero budget is announced, and our zero-valued bid (according to Eq.~(\ref{eq:offered_bid})) is chosen as the winner of the auction, {\it the received packet is  immediately dropped after its reception}. Notice that, since the announced fine is necessarily zero (it cannot exceed the budget and it must be nonnegative), we do not loose any money by not forwarding the received packet. Moreover, by dropping the received packet, we not only frustrate the ``work-for-free'' strategy set up by the auctioneer, but also force it to pay any fine he has agreed upon with any upstream node. In addition, the resulting ``cascade effect'' on fine payment imposed on the auctioneer's upstream nodes may cause them to avoid this node as a relay in future auctions. By doing so, we expect that the more nodes discourage such behavior, the less are the chances of having someone acting in a greedy manner. This rule is a way of signaling greedy nodes that they must change their way of announcing RFBs in order to promote a more cooperative behavior and better sharing of resources and monetary gains.

\section{Lessons Learned \& Discussion}
\label{sec:lessons_learned}

A number of lessons can be drawn from the participation in the MANIAC Challenge 2013, especially regarding the use of of off-the-shelf tablets and/or smartphones in the deployment of ad hoc networks for purposes of infrastructure offloading through auction-based forwarding schemes.

 


\subsection{MAC and Routing Protocols}

As far as the adoption of tablets and/or smartphones for the deployment of ad hoc networks is concerned, a number of issues still need to be addressed in order to have these devices fully functional for a lasting and reliable multi-hop ad hoc operation. Today, practically all tablets and smartphones support the IEEE 802.11 a/b/g/n. However, in spite of supporting WiFi, the Android platform does not officially support its ad hoc mode of operation, and off-the-shelf tablets/smartphones can only operate in infrastructure mode by default. Therefore, as of this writing, there is still the need to resort to command line hacking in order to have the ad hoc mode of operation working on Android-based devices. This is unfortunately different between different devices and ROMS.
This lack of support for ad hoc mode does not only impact the operation at the medium access control (MAC) sub-layer, but also impacts the installation and operation of the routing protocol. As mentioned before, the MANIAC Challenge 2013 has adopted the OLSR as the routing protocol for ad hoc networking. In particular, they have adopted the OLSRd~\cite{olsr} implementation for Android platform. In addition to having to undergo the same installation issues mentioned before, the OLSRd implementation does not provide any enhancements regarding {\it energy efficiency}. OLSR is a proactive, table-driven routing protocol that constantly broadcasts topology and other control information to the entire network. In spite of using a clever idea of choosing a subset of nodes to propagate broadcast information---the multipoint relays (MPRs)---its original design did not target energy efficiency~\cite{taddia.secon.2006,derango.milcom.2008,mahfoudh.ainaw.2008,kunz.iwcmc.2008}. Hence, in order to have a lasting and efficient ad hoc operation for offloading purposes, the energy efficiency of the underlying routing protocol must be urgently addressed. Last, but not least, a major problem that has also been found was incompatible WiFi drivers that led to incorrect behavior between end devices~\cite{ietf-maniac}.

\subsection{The Issue of Timeouts}

In the MANIAC Challenge 2013, timeouts were announced in terms of {\it maximum number of hops to destination}, as opposed to an actual time value. With this definition, it is implicit that the backbone itself translates the {\it maximum time allowed to deliver a packet to its destination} into a {\it maximum number of hops that a packet can traverse across the ad hoc network to its destination}. As a result, our ``tightness'' concept was built around that specific definition. In another possible scenario, the backbone would simply announce an actual time value as the deadline. If that were the case, we would need to modify our strategy accordingly. This means that, instead of computing the ``surplus'' or ``deficit'' of hops through the shortest path to destination from a given node $i$ (tightness function $\Delta_i$ in Section~\ref{sec:strategy}), we would need to find out the {\it remaining time through the path with the shortest delay from a node $i$}. Consequently, a significant modification to our strategy would be needed, since it would require the estimation of delay over different paths. This is not an easy task, and it would require an estimation technique that is simple, fast, and able to deliver accurate results.    

\subsection{Cooperative versus Selfish Behavior}

The issue of cooperation is critical in order to make data offloading feasible with auction-based forwarding schemes in ad hoc networks. Eventually, every node wants to receive {\it some} share of the budget offered by the backbone. Otherwise, there will be no point in participating in the ad hoc network. In fact, it is reasonable to expect that some node(s) may end up receiving more than others as a result of their strategies, but {\it some} fraction of their gains should be shared among everyone (at least in the long run). Therefore, it is key that every participant node perceives some level of fairness and ``profit sharing'' within the network. To accomplish that, announced auctions should be as attractive as possible, so that nodes keep motivated to join and participate in the ad hoc network. On the other hand, whenever possible, nodes should strive to punish those nodes that act in an unfair manner or uncooperatively. One such technique for that is the one we have implemented in our strategy. 

\subsection{Need for Large-Scale Performance Evaluations} 

The MANIAC Challenge 2013 was a very rich experience, and it has put together very different and creative strategies to compete under the proposed offloading architecture. However, because the network comprised 12 tablets only (including a pair of ``dummy'' tablets), and the event lasted a single day, it was not possible to carry out a thorough investigation on scalability issues, for instance. In fact, while the introduction of more nodes would certainly improve geographical coverage and provide higher path redundancy (i.e., fault tolerance), it might as well degrade link quality and network delay due to channel contention and higher interference. Therefore, it would be interesting to observe the behavior of the offloading architecture under significant traffic load and higher degree of interaction. Moreover, apart from the network itself, it also remains unclear how individual strategies would perform under such circumstances. Therefore, there is a clear need for a thorough, large-scale performance evaluation of such auction-based forwarding schemes for data offloading. Such evaluations would be instrumental for assessing the actual feasibility of mobile data offloading through wireles ad hoc networks.


\section{Conclusions}
\label{sec:conclusions}

As demonstrated by the MANIAC Challenge 2013, there are many issues still to be addressed before mobile data offloading through wireless ad hoc networks becomes a viable and functional solution across many devices and mobile operating systems. However, the richness of application scenarios and the creativity of solutions presented at the event suggest that wireless ad hoc networks may help to not only offload the operator's infrastructure, but also to provide smart data pricing schemes and extended wireless coverage. 
      
\section{Acknowledgments}
\label{sec:acknowledgments}

The authors thank the support provided by the University of Brasilia, the Faculty of Technology (UnB), the LATITUDE Laboratory (UnB), Centro de Pesquisa em Arquitetura da Informa\c c\~ao (CPAI), On Telecom, Poupex, Coordena\c{c}\~ao de Aperfei\c{c}oamento de Pessoal de N\'ivel Superior (CAPES), Conselho Nacional de Desenvolvimento Cient\'ifico e Tecnol\'ogico (CNPq), and the wonderful support provided by the MANIAC Challenge 2013 organizers and volunteers. 

\bibliographystyle{IEEE}







\bibliography{references}

\begin{thebibliography}{1}

\bibitem{att}
``Mobile data traffic surpasses voice,''
  http://www.cellular-news.com/story/42543.php.

\bibitem{maniac.challenge.2013}
``{MANIAC} challenge 2013,'' http://2013.maniacchallenge.org/.

\bibitem{clausen2003olsr}
Thomas Clausen, Philippe Jacquet, C{\'e}dric Adjih, Anis Laouiti, Pascale
  Minet, Paul Muhlethaler, Amir Qayyum, Laurent Viennot, et~al.,
\newblock ``Optimized link state routing protocol ({OLSR}),'' Oct 2003,
\newblock RFC 3626.

\bibitem{olsr}
``{OLSRd},'' http://www.olsr.org/.

\bibitem{taddia.secon.2006}
C.~Taddia, A.~Giovanardi, and G.~Mazzini,
\newblock ``Energy efficiency in {OLSR} protocol,''
\newblock in {\em Proc. 3rd Annual IEEE Sensor and Ad Hoc Communications and
  Networks (SECON)}, 2006, vol.~3, pp. 792--796.

\bibitem{derango.milcom.2008}
F.~De~Rango, M.~Fotino, and S.~Marano,
\newblock ``{EE-OLSR}: Energy efficient {OLSR} routing protocol for mobile
  ad-hoc networks,''
\newblock in {\em Proc. IEEE Military Communications Conference (MILCOM)},
  2008, pp. 1--7.

\bibitem{mahfoudh.ainaw.2008}
S.~Mahfoudh and P.~Minet,
\newblock ``An energy efficient routing based on {OLSR} in wireless ad hoc and
  sensor networks,''
\newblock in {\em Proc. International Conference on Advanced Information
  Networking and Applications - Workshops (AINAW)}, 2008, pp. 1253--1259.

\bibitem{kunz.iwcmc.2008}
T.~Kunz,
\newblock ``Energy-efficient variations of {OLSR},''
\newblock in {\em Proc. of 8th International Wireless Communications and Mobile
  Computing Conference}, 2008, pp. 517--522.

\bibitem{ietf-maniac}
Emmanuel Baccelli, Felix Juraschek, Oliver Hahm, Thomas~C. Schmidt, Heiko Will,
  and Matthias Wahlisch,
\newblock ``The {MANIAC} challenge at {IETF} -- student competition makes
  network problem solving fun,''
\newblock {\em The IETF Journal}, vol. 9, no. 2, pp. 27--29, 2013.

\end{thebibliography}
\end{document}